\documentclass[sigconf]{acmart}
\renewcommand\footnotetextcopyrightpermission[1]{} %
\usepackage{amsmath,amssymb,amsfonts}
\usepackage{algorithm2e}
\usepackage{algorithmic}
\usepackage{graphicx}
\usepackage{textcomp}
\usepackage{xcolor}
\def\BibTeX{{\rm B\kern-.05em{\sc i\kern-.025em b}\kern-.08em
    T\kern-.1667em\lower.7ex\hbox{E}\kern-.125emX}}
    
\usepackage{subcaption}
\usepackage{balance}
\usepackage{listings}
\usepackage{url}
\usepackage[most]{tcolorbox}
\usepackage{multirow}

\usepackage{enumitem}

\definecolor{codegreen}{rgb}{0,0.6,0}
\definecolor{codegray}{rgb}{0.5,0.5,0.5}
\definecolor{codepurple}{rgb}{0.58,0,0.82}
\definecolor{backcolour}{rgb}{0.95,0.95,0.92}
\definecolor{lightgray}{gray}{0.9}   
  
\definecolor{yellow}{RGB}{255,255,153}
\definecolor{grey}{RGB}{224,224,224}

\newboolean{showcomments}
\setboolean{showcomments}{true}
\ifthenelse{\boolean{showcomments}}
 { \newcommand{\mynote}[2]{
      \fbox{\bfseries\sffamily\scriptsize#1}
        {\small$\blacktriangleright$\textsf{\emph{#2}}$\blacktriangleleft$}}}
        { \newcommand{\mynote}[2]{}}

\definecolor{DarkOrange}{rgb}{0.8,0.3,0.0} 
\definecolor{DarkCyan}{rgb}{0.0, 0.55, 0.55}
\definecolor{Gray}{gray}{0.9}

\newcommand{\tool}{\textsc{Anchor}\xspace}  %
 
\begin{document}
\author{Pingfan Kong}
\affiliation{\institution{University of Luxembourg}\country{Luxembourg}}
\email{pingfan.kong@uni.lu}

\author{Li Li}\authornote{Corresponding author.}
\affiliation{\institution{Monash University}\country{Australia}}
\email{li.li@monash.edu}

\author{Jun Gao}
\affiliation{\institution{University of Luxembourg}\country{Luxembourg}}
\email{jun.gao@uni.lu}

\author{Timoth\'ee Riom}
\affiliation{\institution{University of Luxembourg}\country{Luxembourg}}
\email{timothee.riom@uni.lu}

\author{Yanjie Zhao}
\affiliation{\institution{Monash University}\country{Australia}}
\email{yanjie.zhao@monash.edu}

\author{Tegawend\'e F. Bissyand\'e}
\affiliation{\institution{University of Luxembourg}\country{Luxembourg}}
\email{tegawende.bissyande@uni.lu}

\author{Jacques Klein}
\affiliation{\institution{University of Luxembourg}\country{Luxembourg}}
\email{jacques.klein@uni.lu}
\title{\tool: Locating Android Framework-specific Crashing Faults}

\begin{abstract}
Android framework-specific app crashes are hard to debug. Indeed, the callback-based event-driven mechanism of Android challenges crash localization techniques that are developed for traditional Java programs. The key challenge stems from the fact that the buggy code location may not even be listed within the stack trace.  For example, our empirical study on 500 framework-specific crashes from an open benchmark has revealed that 37 percent of the crash types are related to bugs that are outside the stack traces. Moreover, Android programs are a mixture of code and extra-code artifacts such as the Manifest file. The fact that any artifact can lead to failures in the app execution creates the need to position the localization target beyond the code realm. In this paper, we propose \tool{}, a two-phase suspicious bug location suggestion tool. \tool{} specializes in finding crash-inducing bugs outside the stack trace. \tool{} is lightweight and source code independent since it only requires the crash message and the apk file to locate the fault. Experimental results, collected via cross-validation and in-the-wild dataset evaluation, show that \tool{} is effective in locating Android framework-specific crashing faults.
\end{abstract}

\settopmatter{printacmref=false}
\setcopyright{none}


\maketitle
\pagestyle{plain} %


\section{Introduction}\label{sec:introduction}

App crashes are a recurrent phenomenon in the Android ecosystem~\cite{wei2016taming}. They generally cause damages to the app reputation and beyond that to the provider's brand~\cite{crashesAndroidDev}. Apps with too many crashes can even be simply uninstalled by annoyed users. They could also receive bad reviews which limit their adoption by new users. 
Too many apps crashes could also be detrimental to specific app markets that do not provide mechanisms to filter out low-quality apps concerning proneness to crash. The challenges of addressing Android app crashes have attracted attention in the research community.

 Fan et al.~\cite{fan2018large} have recently presented insights on their large-scale study on framework-specific exceptions raised by open source apps. In more recent work, Kong et al.~\cite{kong2019-ISSTA-crashes} have proposed an automated approach to mine fix patterns from the evolution of closed-source apps (despite the lack of change tracking systems). Tan et al.~\cite{tan2018repairing} further presented an approach to repair Android crashing apps. 
 A common trait of all these crash-related studies is that the underlying approaches heavily rely on the generated stack traces to identify the fault locations.  
 Although the state of the art is effective for many bugs, they are generally tailored to the generic cases where the stack traces provide relevant information for locating the bug. Unfortunately, there is a fair share of faults whose root causes may remain invisible outside the stack trace.
Wu et al.~\cite{wu2014crashlocator} have already reported this issue in their tentative to locate crashing faults for general-purpose software.
In the realm of Android, the phenomenon where the stack trace may be irrelevant for fault localization is exacerbated by two specificities of Android:

{\em The Android system is supported by a callback-based and event-driven mechanism:} 
Each component in Android has its lifecycle and is managed by a set of callbacks. Every callback serves as a standalone entry point and root to a separate call graph. Yet, existing crash-inducing bug localization techniques for Java such as CrashLocator~\cite{wu2014crashlocator} assume a single entry point to compute certain metrics for the suspiciousness score of different methods. Additionally, since the Android system is event-driven, the invocation sequence to functions and callbacks is affected by non-deterministic user inputs or system events, making the stack trace unreliable for quick analyses. 
	
{\em The Android app package includes both code and resources that together form the program:}
Android apps are more than just code. They are combinations of Java/Kotlin code, XML files, and resources (such as images and databases). 
Apps provide extensions to the Android Operating System (OS), which directly analyses XML files from the app to map callback functions, which the OS must trigger to exploit functionalities in the apps. 
Therefore, an error by developers within an XML document can eventually lead to a runtime crash. Similarly, it is important to note that crashes can occur due to other concerns such as the arrangements of app resources, use of deprecated APIs, omissions in permission requests, etc. 
Typical such errors, which occur outside of code pointed to by stack traces, will cause either developers or Automatic Program Repair (APR) tools (e.g.,~\cite{tan2018repairing}) to pointlessly devote time in attempting to fix the code.
 
{\bf This paper.} Our work aims at informing the research community on the acute challenges of debugging framework-specific crashes. To that end, we propose to perform an empirical study that investigates the share of crashes that cannot be located by current localization approaches. 
Following this study, we present a new approach to locate faults, aiming at covering different categories of root cause locations. Overall, we make the following contributions: 
\begin{itemize} 
	\item We present the results of an empirical study performed on a set of 500 app crashes retrieved from the ReCBench dataset~\cite{kong2019-ISSTA-crashes}. A key finding in this study is that we were able to identify that 37\% crash 
		root causes are associated with crash cases where the stack trace is not directly relevant for fault localization.
	\item We propose \tool, a tool-supported approach for locating crashing faults. \tool unfolds in two phases and eventually yields a ranked list of location candidates. The first phase applies a classification algorithm to categorize each new crash into a specific category. Depending on this category, a dedicated localization algorithm is developed in the second phase. \tool currently implements 3 localization algorithms that eventually generate a ranked list of buggy methods (when the bug is in the code) or resource types (when it is outside of code).
	\item We performed 5-fold cross-validation on the 500 crash cases to assess the effectiveness of \tool{} in placing the crashing fault location in the top of its ranked list of suggestions. \tool{} exhibited an overall MRR (Mean Reciprocal Rank) metric value of 0.85. An analysis of the open dataset of crashed open-source Android apps further shows that our method scales to new app crashes.
\end{itemize}

The rest of this paper is organized as follows. Section~\ref{sec:background} introduces background details on Android app crashes and callback-based event-driven mechanisms. Section~\ref{sec:motivation} revisits the motivating example by the previous work~\cite{tan2018repairing} and demonstrates why research in crash localization has standing challenges. Section~\ref{sec:empirical} discusses the findings of our empirical study and explores the insights that can be leveraged for a new approach. Section~\ref{sec:implementation} presents \tool{}. We describe experimental setup in Section~\ref{sec:setup} and approach evaluation in Section~\ref{sec:evaluation}. We bring further discussion in Section~\ref{sec:discussion}. Threats to validity are acknowledged in Section~\ref{sec:validity} and related work is presented in Section~\ref{sec:related}. Finally, Section~\ref{sec:conclusions} concludes the paper.

\section{Background}\label{sec:background}
In this section, we introduce the important concepts related to this paper.

\subsection{Android App Crash Stack Trace}
\begin{figure}[!h]
	\centering
	\includegraphics[width=1\linewidth]{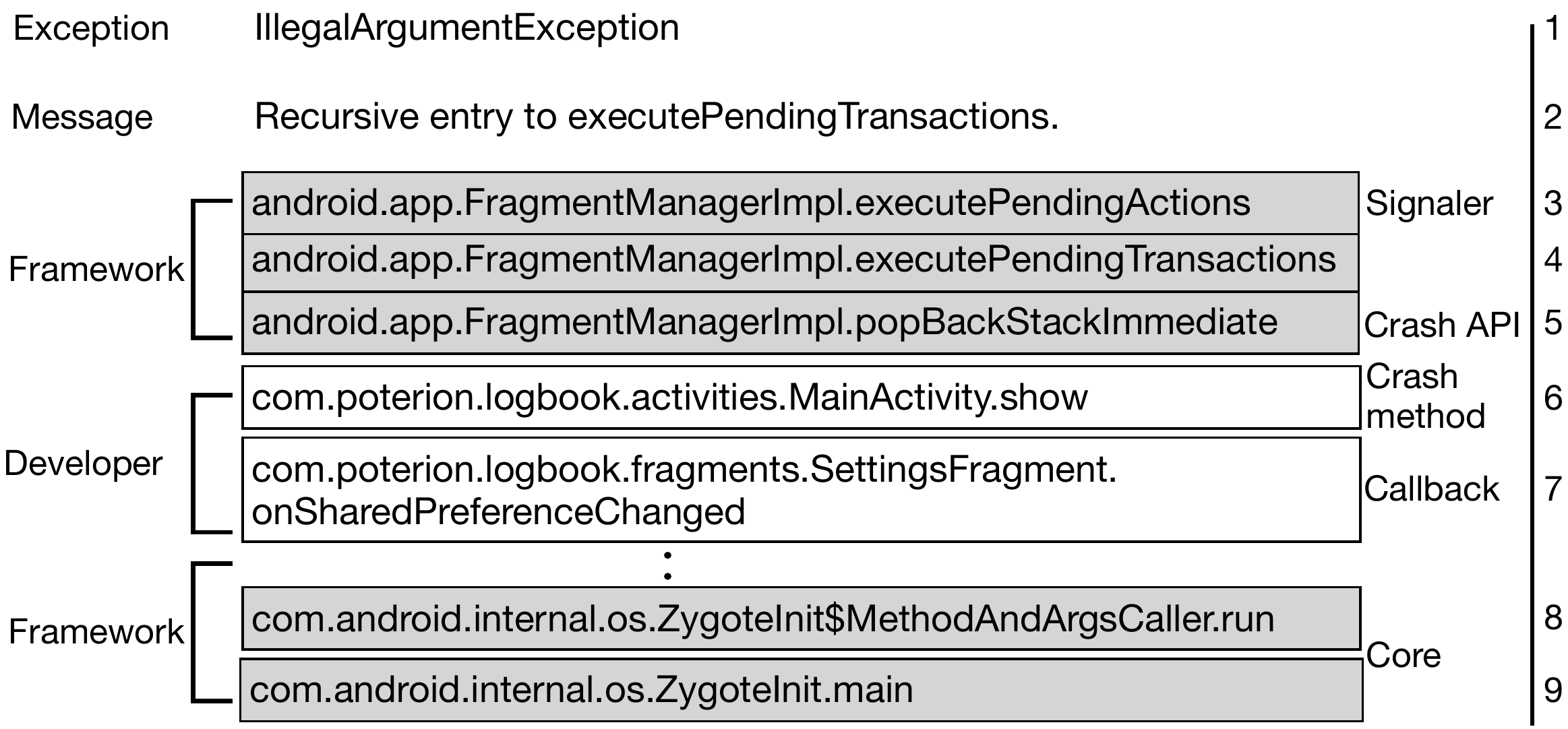}
	\caption{Crash Stack Trace of app Sailer's Log Book.}
	\label{fig:sandwich}
\end{figure}

Like all Java\footnote{Kotlin has also been widely used in recent years as an alternative for Android app development, it is designed to fully interoperate with Java.} based software, when Android apps crash, they can dump execution traces which include the exception being thrown, a crash message, and most importantly, a stack trace of a callee-caller chain starting from the \emph{Signaler}, i.e., the method that initially constructed and threw the exception object. 
Figure~\ref{fig:sandwich} is an example of stack trace for the crash of the app \emph{Sailer's Log Book}. This app helps sailors to keep their logbook accurate and up-to-date. On the first line, the exception \emph{IllegalArgumentException} is thrown. On the second line, the log system reports message "recursive entry to executePendingTransactions". Starting from the third line, the \emph{Signaler} of the stack trace is listed: it is this framework method that instantiates the exception type, composes the log message and throws it to its caller to handle. On Lines 4-5 that are also marked in grey, there are other two methods that continue to pass on the exception. Line 5 holds the \emph{API}, which is the only framework method in this stack trace that is visible to the developer. Since the crash happens directly due to invocation to it, we call it the \emph{Crash API}. Line 6 is the developer method that invoked this API. Line 7 is the developer implementation of the callback, inherited from the superclass of the Android framework. Android framework decides, based on certain conditions and system/user events, when to invoke this method, and what parameter values to pass in. Lines 8-9 are part of the Android framework core that is, again, not accessible to developers.

The crash stack trace is often the first thing that developers want to examine when a crash is reported~\cite{kim2011crashes}. Even when it is not given, developers would reproduce and retrieve them. Intuitively, the crash arises from mistakes in the developer methods, e.g., Lines 6-7 in Figure~\ref{fig:sandwich}. Particularly, the \emph{Crash method} that directly invoked the \emph{Crash API}. Our empirical study in Section~\ref{sec:empirical} shows that this intuition is correct, that 63\% of the total crash types are in the stack trace. However, in the rest of this section, we will introduce the specialty of Android that may lead to the rest 37\%.

\subsection{Callback-based and Event-driven Mechanism}\label{subsec:callback-based}
Unlike traditional Java programs, Android apps have multiple entry points. Each entry point is a callback method (e.g., Line 7 in Figure~\ref{fig:sandwich}), which is declared in one of the Android framework component classes, inherited by the developer defined subclass, and maybe overridden by the developer. The Android framework core, based on the user inputs and system environments, decides when to invoke the callbacks and what parameter values to pass in. Each callback is standalone, and in general Android does not encourage developers to invoke those callbacks from their self-defined methods, unless these methods are callbacks overriding and invoking their super. As a result, existing static call graph based fault localization techniques~\cite{wu2014crashlocator} for Java programs can not be simply reused, since they assume single entry points and need to compute weighing scores based on the graph. There are, however, works~\cite{yang2015static,octeau2015composite} that have invented methods to track the control flows or data flows and tried to build the callback connections.  These proposed approaches are either computationally expensive or confined in limited number of component classes, and does not scale to all scenarios. Other approaches like~\cite{li2016droidra} or~\cite{arzt2014flowdroid} create a dummy main to invoke all callbacks in order to reuse Java based analysis tools, but this method discarded the relation among callbacks, which is crucial to estimate the possibility of a method containing the real bug.

\begin{figure}[!h]
	\centering
	\includegraphics[width=0.95\linewidth]{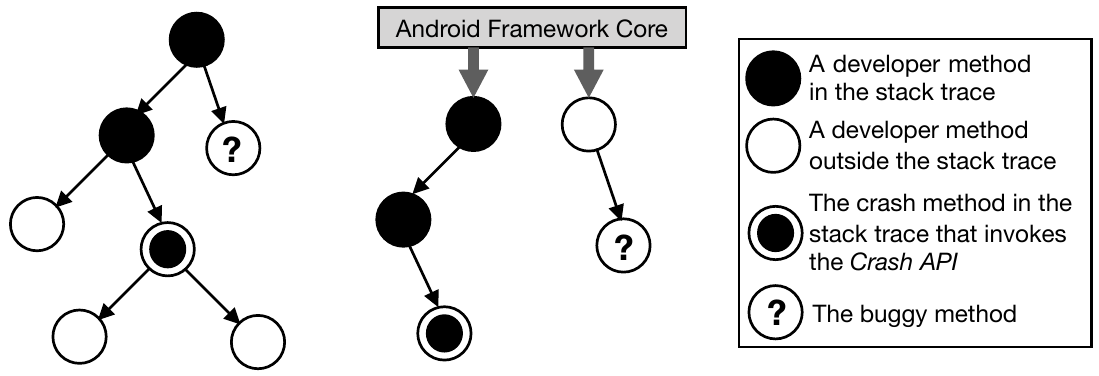}
	\caption{Call Graph Comparison between General Java Program (left) and Android App (right), inspired from~\cite{wu2014crashlocator}}
	\label{fig:callback_model}
\end{figure}

Figure~\ref{fig:callback_model} examplifies the difference of call graphs between general Java program (left) and Android app (right). The filled circles represent the developer methods in the stack trace, while the non-filled circles represent developer methods outside the stack trace. The partially filled circles represent the \emph{Crash method} that invokes the \emph{Crash API}. 
Generally, the buggy method is the \emph{Crash method}. However, as shown in our empirical study, it appears that the buggy method (the circle filled with question mark in Figure~\ref{fig:callback_model}) is not connected to the \emph{Crash method}. 
A traditional Java program static call graph based approach such as CrashLocator~\cite{wu2014crashlocator} will be able to locate this buggy method only if the buggy method is "close enough" to the generated call graph (roughly speaking they generate an extended call graph leveraging the stack trace). 
However, on the right, in the case of Android apps, the buggy method could be in a separate call graph
because of callback methods that are invoked by the Android framework. Such cases will be missed by approaches such as CrashLocator~\cite{wu2014crashlocator} that only detects buggy methods captured by its extended call graph, but does not consider callback methods.

\subsection{Android APK File Format}\label{subsec:apk_format}
Android apps are distributed in a package file format with extension ".apk". It is a compressed folder containing code, resources, assets, certificates, and manifest file. All of these files are crucial to the expected good functioning of the apps. Therefore, some crashes may be induced when there are problems with these files.

\subsubsection{Android Manifest File}
Every app project needs to have an AndroidManifest.xml file at the root of the project source set~\cite{androidmanifest}. Along with the package name and components of the app, this file also declares the permissions that the apps needs, as well as the hardware and software features that the app requires. 

\subsubsection{Android Component Layout Description File}\label{subsubsec:layout}
Android component layout description files are also crucial to the execution of the app. E.g., Listing~\ref{listing:layout} is the layout file of the main Activity of an Android app \emph{Transistor}. In this file, a child fragment is defined and described. 
The attribute \emph{android:id} defines the layout file name to be inflated for the fragment, the attribute \emph{android:name} gives the full name of the user defined Fragment class. When the main Activity is being created, the Android framework scans this layout file, and invokes a series of relevant callbacks on this Fragment to draw it along with the main Activity.

    \begin{lstlisting}[caption={Main Activity Layout File of app Transistor.}, language=xml,breaklines=true,linewidth={\linewidth},basicstyle=\footnotesize\ttfamily,label=listing:layout, float=h]
<?xml version="1.0" encoding="utf-8"?>
<fragment xmlns:android="http://schemas.android.com/apk/res/android"
    xmlns:tools="http://schemas.android.com/tools"
    android:id="@+id/fragment_main"
    android:name="org.y20k.transistor.MainActivityFragment"
    android:layout_width="match_parent"
    android:layout_height="match_parent"
    tools:layout="@layout/fragment_main" />
\end{lstlisting}

\section{Motivating Example}\label{sec:motivation}

\begin{figure}[!h]
	\centering
	\includegraphics[width=0.9\linewidth]{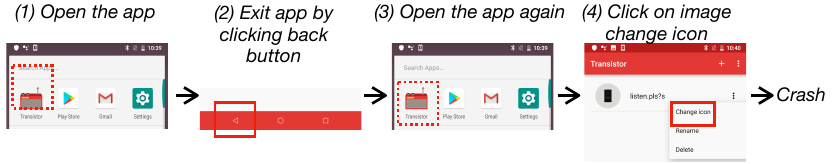}
	\caption{Crash of Transistor.}
	\label{fig:motivating_example}
\end{figure}

We further illustrate the challenges of locating faults outside Android app stack traces by revisiting an example that was used to motivate a previous work on Android app crash automatic repairing by Tan et al.~\cite{tan2018repairing}. 
{\em Transistor}\footnote{https://github.com/y20k/transistor/issues/21} is a popular online radio streaming app. We showed its partial resources in Section~\ref{subsubsec:layout}. However, it was reported that following the input sequence in Figure~\ref{fig:motivating_example}, the app will crash.

\begin{lstlisting}[caption={Crash Message of \emph{Transistor}.},language=xml,numbers=left,breaklines=true,label=listing:stack_trace]
java.lang.IllegalStateException: 
MainActivityFragment{e7db358} not attached to Activity
at ...MainActivityFragment.startActivityForResult(Fragment.java:925) (Crash API)
at ...agment.selectFromImagePicker(MainActivityFragment.java:482) (Crash method)
at ...k.transistor.MainActivityFragment.access$500(MainActivityFragment.java:58)
at ...transistor.MainActivityFragment$6.onReceive(MainActivityFragment.java:415)
\end{lstlisting}

The crash message filtered out from logcat is shown in Listing~\ref{listing:stack_trace}. 
It appears that invoking the \emph{startActivityForResult} \emph{API} on the {\tt MainActivityFragment} (line 3) throws an unhandled \emph{IllegalStateException} (line 1), and the Android system reports that the fragment is not attached to the hosting activity (line 2). 
By inspecting the source code of Android framework of the \emph{Crashed API} (line 3), we see that the \emph{startActivityForResult} method of the fragment instance attempts to invoke its context's (i.e., its host {\tt Activity}'s) \emph{API} with the same name \emph{startActivityForResult}. This invocation is guarded by an if-clause, which checks whether the fragment is still attached to the host {\tt Activity}. If not, however, the \emph{IllegalStateException} will be thrown.

\begin{lstlisting}[caption={Fix from Tan et al.}, language=Java,breaklines=true,linewidth={\linewidth},basicstyle=\footnotesize\ttfamily,label=listing:fix_fan, float=h]
new BroadcastReceiver(){
	onReceive(...){	...
+		if(getActivity()!=null) 
  		startActivityForResult(pickImageIntent,REQUEST_LOAD_IMAGE);}}
\end{lstlisting}

Biased by the assumption that the fault should only be in the developer methods in the stack trace (lines 4-6), Tan et al.~\cite{tan2018repairing} proposed to amend the \emph{Crash method} (line 4). Listing~\ref{listing:fix_fan} shows their fix. Their fix applies a checker on invocation to \emph{startActivityForResult}, which will not be executed if value of \emph{getActivity} is null (i.e., when the fragment is no longer attached to its hosting {\tt Activity}). As a result, the app avoids crashing.
This fix indeed prevents the exception. However, it is not explainable: applying the checker not only prevents the crash, but it should also prevent opening the \emph{SelectImageActivity} as designed for. Due to this paradox, we have a good reason to suspect that the true bug location is still hidden.

{\em Transistor}'s developer, who is also dedicated in debugging in the stack trace, proposed a fix on her/his own in Listing~\ref{listing:fix_developer}. Realizing that the \texttt{Fragment} lost its reference to the host {\tt Activity}. The developer declared a variable \texttt{mActivity} to hold the reference. Then in the \emph{Crash method} (line 4 in Listing~\ref{listing:stack_trace}), she/he switched the invocation of the \emph{startActivityForResult} \emph{API} from \texttt{Fragment} to \texttt{mActivity}.

\begin{lstlisting}[caption={Fix from Developer.}, language=Java,linewidth={\linewidth},breaklines=true,basicstyle=\footnotesize\ttfamily,label=listing:fix_developer, float=h]
+	mActivity = getActivity();	...
new BroadcastReceiver(){
	onReceive(...){	...
-		startActivityForResult(pickImageIntent,REQUEST_LOAD_IMAGE);
+		mActivity.startActivityForResult(pickImageIntent,REQUEST_LOAD_IMAGE);}}
\end{lstlisting}

This fix also bypassed the crash, but it causes regression. After the final step in Figure~\ref{fig:motivating_example}, if the user clicks on the back button two more times, the app should have first returned to the \texttt{MainActivity}, then back to the home screen. Instead, it opens another \emph{SelectImageActivity}. In the issue tracking, the developer admits that she/he had no idea of how to fix it. While after several months, the bug "fixed" itself, which she/he described as "scary". Even Tan et al. failed to explain the cause of this regression.

Based on the understanding of Android' callback-based mechanism introduced in Section~\ref{subsec:callback-based}, we suspect that the bug may not exist in the stack trace. We confirmed our fix shown in Listing~\ref{listing:fix_us}. This fix is reported to the developer and we received positive feedback in the issue tracking, as can be verified in {\em Transistor}'s respository given above. 

\begin{lstlisting}[caption={Fix Inspired by this Article.}, language=Java,breaklines=true,linewidth={\linewidth},basicstyle=\footnotesize\ttfamily,label=listing:fix_us, float=h]
MainActivityFragment extends Fragment{ 
 	onDestroy(){
+ 	   super.onDestroy();
+		LocalBroadcastManager.getInstance(mApplication).unregisterReceiver(imageChangeRequestReceiver,imageChangeRequesIntentFilter);}}
\end{lstlisting}

We broaden the search for bug outside the stack trace. Noticing the crash originated from the \emph{onReceive} callback (cf. line 6 in Listing~\ref{listing:stack_trace}), we examine the lifecycle of this \texttt{BroadcastReceiver} object. We found that it is registered in the \emph{onCreate} callback of \texttt{MainActivityFragment}, but never unregistered in its counterpart callback \emph{onDestroy}. As a result, after Step 2 (cf. Figure~\ref{fig:motivating_example}), the \texttt{BroadcastReceiver} and its host \texttt{MainActivityFragment} are leaked in the memory. In Step 4, the callbacks of the leaked objects are stealthily invoked by the Android framework and eventually caused the \emph{IllegalStateException}. Knowing the true cause of the crash, it is not difficult to explain the paradox of Tan et al.'s fix and the regression caused by the developer's fix. However, given the page limit, we put detailed reasoning online at \underline{https://anchor-locator.github.io}.

\begin{tcolorbox}[breakable,boxsep=5pt,arc=8pt,sharp corners=downhill,top=0mm,left=0mm, right=0mm, bottom=0mm]
{\bf Hint:} The fault locations in Android apps may: (1) Be outside the stack trace; (2) Be even outside the call graph extended from the stack trace; (3) Not even ``exist'' in the code, i.e., they are inherited methods without visible code snippets. Locating such faults may require tremendous efforts. Fixes based on incorrect localization may even cause regression.
\end{tcolorbox}

\section{Empirical Study on Fault Locations}\label{sec:empirical}

In this section, we present the results of an empirical study that we performed on a set of 500 app crashes retrieved from the ReCBench dataset~\cite{kong2019-ISSTA-crashes}. This study aims at assessing to what extent the locations of crashing faults reside outside the stack trace. 

\subsection{Dataset Construction}\label{subsec:dataset_construction}
We extract our dataset from ReCBench, an open dataset proposed by Kong et al.~\cite{kong2019-ISSTA-crashes} in 2019. 
ReCBench has been built by running hundreds of thousands of Android apps downloaded from various well-known Android markets~\cite{allix_androzoo:_2016,li2017androzoo++}. 
In addition to a collection of crashed Android apps focusing on framework-specific crashes\footnote{Android framework methods are not visible or understandable to general developers, hence greater challenge is acknowledged for locating framework-specific crashes compared to developer-written methods.~\cite{fan2018large,kong2019-ISSTA-crashes}}, ReCBench offers the possibility to collect crash log messages and scripts to reproduce the crashes. 
Today, ReCBench contains more than 1000 crashed apps (still growing). For our empirical study, we focus on crashed apps for which: 
\begin{itemize}
	\item First, the stack trace of the crash contains at least one developer method. This is a requirement to be able to start an exploration process to find the crash root cause.  %
	\item Second, since we specifically target the crashes induced by Android \emph{API}s, the \emph{Signaler} must be Android-specific.
\end{itemize}
After applying these two rules, we randomly selected $500$ crashed apps from the remaining apps. The dataset is publicly accessible at \underline{https://github.com/anchor-locator/anchor}.

\subsection{Ground Truth \& Results}
We manually inspect all the 500 crashed apps to understand the reason behind the crashes and to create our ground truth. We perform this manual inspection by leveraging the CodeInspect~\cite{codeinspect} tool, following same protocols discussed in~\cite{fan2018large}.
Each of the crashed apps has been categorized into one of the following categories:
\begin{itemize}
	\item Category A groups the crashed apps for which the buggy method (i.e., the root cause) is one of the developer methods present in the stack trace;
	\item Category B groups the crashed apps for which the buggy method is not present in the stack trace, but still in the code. 
	\item Category C groups the crashed apps for which the crash arises from non-code reasons. 
\end{itemize}

The above partition is one out of many alternatives, e.g., one can also separate bugs based on whether they are concurrent~\cite{wang2018aatt+,bielik2015scalable,li2016effectively,tang2016generating,maiya2014race}. We later show in Section~\ref{subsec:localization} how this partition helps with building our localization tool.
Table~\ref{tab:loc_empirical} summarizes the outcome of the empirical study. 
It appears that for 89 (49+40) crashed apps (representing 18\% of the total cases), the crashing fault location is not in any of the developer methods present in the stack trace. The respective numbers of Categories B and C are close, with 49 cases in Category B and 40 cases in Category C. 
To further investigate how many types of crashes occur for each category, we use a method similar to the one described in~\cite{kong2019-ISSTA-crashes}: we group crashes from a given category into  \emph{buckets}. Specifically, if two crash cases have identical framework crash sub-trace, they will be put into the same bucket. 
The last two columns in Table~\ref{tab:loc_empirical} present the number of buckets per category. Overall, there are 105 types of crashes (i.e., buckets) in the dataset. 
The percentage of types of crashes in Categories B and C are 16\% and 21\%, respectively. In total, there are 37\% of buckets whose buggy reasons are not shown in the stack traces. Each unique framework crash sub-trace suggests a unique type of crash-inducing bug.  Therefore, considering crash types encountered per the same number of cases (buckets\#/case\#) in each category, more debugging effort will be needed for Categories B and C than in Category A.

\begin{table}[!h]
\centering
\caption{Categories of Fault Locations in Android apps}
\label{tab:loc_empirical}
\resizebox{0.9\linewidth}{!}{
\begin{tabular}{c |c | c |c |c | c | c }
\hline
Category     & stack trace & code & case\# & percent & bucket\# & percent \\
\hline
A  &in & 	yes & 411 &  82\%  & 66 & 63\% \\
B & out & yes & 49 & \textbf{10\%}  & 17 & \textbf{16\%}\\
C & out & no & 40 &\textbf{8\%}  & 22 & \textbf{21\%}\\	
\hline
Total & - & - & 500 & 100\% & 105 & 100\%\\
\hline
\end{tabular}
}
\end{table}

\begin{tcolorbox}[breakable,boxsep=5pt,arc=8pt,sharp corners=downhill,top=0mm,left=0mm, right=0mm, bottom=0mm]
{\bf Hint:} 18\% of the crashes are due to bugs for which the location is outside the stack trace. A significant number of root causes (buckets), i.e., 37\% (16\%+21\%), are associated with cases where the stack trace is not directly relevant for localization.
In even 21\% of the cases, the root causes are not located in the code. 
		
\end{tcolorbox}

We now detail each category in the rest of this Section.

\subsection{Category A: in Stack Trace}
Category A includes all crash cases whose bugs reside in one of the developer methods present in the stack trace. 
Most crashes in our dataset fall into this category. 
It is expected that by default, developers start their debugging process with the methods present in the stack trace~\cite{jiang2010debugging,schroter2010stack,sinha2009fault,indi2016use}. 
The automatic crash-inducing bug repairing tool named Droix~\cite{tan2018repairing} implements a locater, by assuming that the \emph{Crash method} is the bug location in all scenarios. 
However, we also notice that the true crashing fault may reside in other developer methods, in particular when moving downward in the stack trace. An example of such a case is when the caller methods pass an incorrect instance to the crashed developer methods. Generally, much less effort is needed in locating faults in this category. Since the number of suspected methods is limited and their names are already known. Therefore they are not the focus of this paper.

 \subsection{Category B: out of Stack Trace, in the Code}\label{subsec:categoryB_empricial}
It has drawn attention to researchers that Java program crashes can be caused by methods that are not listed in stack traces. Approaches like CrashLocator~\cite{wu2014crashlocator} broadens the search for such faulty methods in extended call graphs from stack traces. We demonstrate in the rest of this section why this broadened search is not enough for Android apps. There are in total 49 cases in this category, each crashed from wrongly handling a framework \emph{API}. Based on the type of the framework \emph{API} (call-in or callback), we further categorize them into two sub-categories: (1) Misused Call-In APIs and (2) Non-Overridden Callback APIs.%

 \subsubsection{Type 1: Misused Call-In APIs (44 cases out of 49)}
 This first type groups crashing faults caused by the misuse of call-in APIs. This means that the bug leading to a crash is due to a buggy explicit invocation of an API from a developer method.
 Moreover, this invocation is often performed from another implemented callback, other than the callback in the stack trace. 
 Since both callback methods are triggered by the framework, it is unlikely that an extended call graph can cover such methods (cf. Figure~\ref{fig:callback_model}).

   \begin{lstlisting}[caption={Bug Explanation to app Geography Learning.}, language=Java,breaklines=true,linewidth={\linewidth},basicstyle=\footnotesize\ttfamily,label=listing:mis-use, float=h]
public class MainActivity extends Activity{
	onCreate(...){
 	try{bindService(intent,serviceConnection,integer);/*Bug Location*/
 	}...}...
 	onDestroy(){unbindService(serviceConnection);/*Crash location*/}}
\end{lstlisting}
Listing~\ref{listing:mis-use} gives a concrete example. 
This example is extracted from an app named \emph{Geography Learning} which helps users to remember geography knowledge in a quiz game format. 
When the \emph{MainActivity}\footnote{The \emph{Main Activity} of an app is the first screen shown to the user when launched.} of this app is launched,  
the callback method \texttt{onCreate} is automatically triggered by the Android framework. Then, this \texttt{onCreate} method invokes the \texttt{bindService} API to bind to \emph{Service}. \emph{Service} is one of the four basic components of Android, and wrongly handling of \emph{Service} is not uncommon~\cite{song2019servdroid} in Android app development. %
When the user exits the \emph{MainActivity}, the Android Framework invokes the \emph{onDestroy} callback method and tries to unbind the \emph{Service} bound in the \texttt{onCreate} method. 
Thereafter, the app crashes with the exception type \emph{IllegalArgumentException}. 
Analysing the message which says: \emph{"Service not registered: com.yamlearning. geographylearning.e.a.e@29f6021}", we understand that the \emph{Service} has not been bound. 
In the method body of the overridden \emph{onCreate} callback, we found that the invocation to API \emph{bindService} was misused. 
Indeed, \emph{bindService} is surrounded by a try-catch clause, and another statement preceding this API invocation threw an exception which redirects the execution flow to the catch block, skipping the invocation to \emph{bindService}.

Out of a total of 49 cases in Category B, 44 falls into this sub-category.

 \subsubsection{Type 2: Non-Overridden Callback APIs (5 cases out of 49)}
This second type includes crashes caused by non-overridden callback APIs. Callbacks, or call-afters, are APIs that are invoked when the Android framework decides so, based on certain system environment change and/or user actions. Callbacks are inherited, when developers define classes that are subclassing Android base component classes. Developers are often required to override certain, although not all, callback APIs. Forgetting to handle these callbacks may cause apps to crash immediately. Moreover, these crashes may often seem flaky, since its reproduction requires re-establishing the same system environments and/or repeating user action sequences. Existing Java crash locators fail to spot such bugs with two reasons: (1) These callback APIs are not in the extended call graphs of stack traces; (2) The method snippets in developer-defined codes do not exist, so are easily missed.

 Listing~\ref{listing:missing} shows an example of this crash type. 
 The app \emph{Fengshui Master} is a Chinese fortune teller app. 
 The app crashes when trying to get a reference to the writable database. 
 However, when the app crashes, the exception \emph{SQLiteDatabaseException} is triggered with a message claiming \emph{"not able to downgrade database from version 19 to 17"}. 
 
  \begin{lstlisting}[caption={Bug Explanation to Android app Fenshui Master.}, language=Java,breaklines=true,linewidth={\linewidth},basicstyle=\footnotesize\ttfamily,label=listing:missing, float=h]
public class com.divination1518.f.s{
	a(..){sqliteOpenHelper.getWritableDatabase();/*Crash location*/}}
public class com.divination1518.g.p extends SQLiteOpenHelper{ ...
+ onDowngrade(..){...}/*Bug Location*/}
\end{lstlisting}

 According to the Android documentation, the app developer needs to implement the callback method \emph{onDowngrade} in the self-defined subclass of \emph{SQLiteOpenHelper}. This callback method will be invoked when the database in storage has a higher version than that of the system distribution. 
Failing to override this callback API immediately crashes the app. 
 Note that the motivating example (cf. Section~\ref{sec:motivation}) also falls into this sub-category. 
 Given the stealthiness of this fault type, it is particularly difficult, even for a human developer, to spot the bug reason without being very familiar with the Android official documentation. Out of a total of 49 cases in Category B, 5 falls into this sub-category. 
 
Note that we use $api_h$ to denote the wrongly handled API (call-in API or callback API) for cases of Category B. This denotation is later needed for Section~\ref{subsubsec:cateB}.

  \subsection{Category C: out of Stack Trace, out of Code}
As introduced in Section~\ref{subsec:apk_format}, except code, an Android apk also contains resources, assets, certificate, and manifest. They are critical to the functioning of the app. As a result, mistakes in those files may also cause crashes. 
Table~\ref{fig:pie} gives a summary of the buggy locations outside of code.
As illustrated, eleven cases of crashes originate from the \emph{Manifest.xml} file. 
Most cases in this type are because the permissions are not properly declared in the manifest. 
Resources include specifically files with ".xml" extension (excluding the \emph{Manifest.xml} file). 
An Android app uses these resource files to store the layout and pieces of information like string values.
If the required resource is missing or wrong, then the app will crash. 
Assets are the large files, like fonts, images, bitmaps. Assets should be put in the correct directory. If the Android framework is not able to find them and it will crash the app.

\begin{table}[!h]
\centering
\caption{Crash Causes of Categorie C}
\label{fig:pie}
\resizebox{0.9\linewidth}{!}{
\begin{tabular}{c |c | c |c |c | c  }
\hline
Sub-Category     & Manifest & Hardware & Asset & Resource & Firmware \\
\hline
Cases &	11	&	5	&	4	&	2	&	18\\
\hline
\end{tabular}
}
\end{table}

Aside from the files inside the apk, some constraints put forward by the device's hardware and firmware, i.e., the framework may also cause the app to crash. 
For example, the Android billing service can only be tested on real devices, if, however, tested on emulators, the app crashes~\cite{androidbilling}. 
Also, since Android is quickly updated with new features and designs, old apps may crash on newly distributed devices, due to reasons like deprecated APIs and new security protocols established by the framework. 
Developers should generally redesign the relevant functionalities, therefore no single buggy location can be decided.

\section{Ranking Suspicious Locations}\label{sec:implementation}
To help developers identify the true fault locations when debugging app crashes, including faults that reside outside the stack traces, we propose \tool{}. 
\tool{} is a fault location recommendation system based on a two-phase approach. 
In the first phase, \tool{} categorizes a given crash in one of the three categories (A, B, or C) with a classification system. 
Then, in the second phase, according to the decided category, \tool{} each adopts a unique algorithm to suggest a rank of locations that are suspected to contain the true faults.
The rest of this section describes Phase 1 and Phase 2 in more detail.

\subsection{Phase 1: Categorization}\label{subsec:categorization}
The first phase aims at assigning a new crash to one of the three categories (A, B, or C). We use the Na\"ive Bayes algorithm~\cite{rish2001empirical} for the categorization. Na\"ive Bayes is one of the most effective and competitive algorithms in text-based classification. It is widely used for spam detection~\cite{metsis2006spam,yang2006approach}, weather forecasting~\cite{walton2010predicting}, etc. It is especially suitable in the scenario when the training set does not contain a large number of records~\cite{huang2011naive}, e.g., our empirical dataset contains merely 500 manually constructed records. 

To construct a vector for each crash record, we feature words extracted from the crash message. The value of each feature dimension is binary, indicating whether a word exists or not in the message.
More specifically, we extract three parts from the crash message: (1) The exception type, which is a Java class (e.g., \texttt{IllegalArgumentException}); (2) The exception message, which briefly describes the reason of the crash, e.g., line 2 in Figure~\ref{fig:sandwich}; (3) The top framework stack frames, each being a Java method, e.g., lines 3-5 in Figure~\ref{fig:sandwich}. For (1) and (3), we use ``.'' as the word separator, for (2), we use space as the separator. 
To avoid overfitting and to save computing resources, we do not need the entirety of the vocabulary to build the vector. In Section~\ref{subsec:feature}, we further discuss how many words are necessary.

With this categorization system, each new crash will firstly be categorized as a type of "A", "B" or "C" before being processed in Phase 2.

\subsection{Phase 2: Localization}\label{subsec:localization}
The goal of this phase is to provide a rank of potential bug locations (in descending order of suspiciousness), in the form of developer methods when the bug is in the code (i.e., Categories A and B) and of sub-categories when the bug is not in the code (i.e., Category C). 
Before presenting in the following sub-sections 3 standalone algorithms, one for each category, we explain how we compute a similarity score between two crashes.  
This similarity score is used in both  Categories B and C localization algorithms.

\textbf{Similarity between two Crashes:}
We quantify the similarity between two crashes $C_1$ and $C_2$ by computing the similarity between their crash messages as presented in Equation~\ref{eq:vote}: 

\begin{equation}\label{eq:vote}
Sim_{C_1,C_2} = Edit\_Sim(seq_{C_1},seq_{C_2})
\end{equation}

$seq$ is the sequence of framework stack frames in a crash message, e.g., lines 3-5 in Figure~\ref{fig:sandwich}. 
$Sim_{C_1,C_2}$ is then computed as the \emph{Edit Similarity}~\cite{qin2011efficient} between $seq_{C_1}$  and $seq_{C_2}$. The intuition here is that when two crashes share similar bug reasons, their $seq$ tends to be similar, if not identical.

\subsubsection{Category A: In Stack Trace}\label{subsubsec:category_A}
Since the crash is assigned to Category A, it indicates that the buggy method is one of the developer methods in the stack trace. We inherit the intuition from~\cite{tan2018repairing}, that if the developer method is closer to the \emph{Crash API} in the stack trace, there is a higher chance that it contains the true fault. Therefore, we can obtain the rank without changing the order of the developer methods in the stack trace. For example, in Figure~\ref{fig:sandwich}, methods on line 6 and line 7 are respectively ranked first and second.

\subsubsection{Category B: Out of Stack Trace, in the Code}\label{subsubsec:cateB}
When the crash is classified into Category B, it indicates that the buggy developer method is not in the stack trace, but still in the code. 
As discussed in Section~\ref{subsec:categoryB_empricial}, the buggy method should either be a developer method that misused a call-in API, or a callback API that has not been overridden. In the remainder of this section, we will note $api_h$ this API (call-in API or callback API) that has been wrongly handled (cf. Section~\ref{subsec:categoryB_empricial}). %
To infer a ranked list of potentially buggy methods, we propose Algorithm~\ref{algo:localizationB}. %
The overall idea is, starting from each developer method in the stack trace, in addition to examining the developer methods (1) in the extend call graph, we also examine those that either (2) control the Android components' lifecycles, or 
 (3) are involved in the data flow of the crash. 
The computation of the suspiciousness score follows the same intuition as explained in Section 5.2.1.

\begin{algorithm}
\caption{Localization Algorithm for Category B}
\label{algo:localizationB}
    \SetAlgoLined
\KwData{$crash$: the crash to resolve}
\KwData{$ST$: List of developer methods in stack trace of $crash$}
\KwData{$api_h$: \emph{Wrongly handled API}}
\KwResult{R: Rank of suspicious developer methods}
  \begin{algorithmic}[1]
  	\STATE $S$ $\leftarrow$ Developer methods that invoke $api_h$\;
   \FOR{$sf$ $\in$ $ST$}
		\IF{$api_h$ type ``call-in''}
			\FOR{$s$ $\in$ $S$}
				\FOR{$am$ $\in$ $AM$}
					\IF{$s$ links $am$}
			    		\STATE $s.score+=\frac{1}{d	}$
			    	\ENDIF
				\ENDFOR
			\ENDFOR
			\STATE $R$ $\leftarrow$ $S.sort()$
		\ELSIF{$api_h$ type ``callback''}
			\FOR{$nc$ $\in$ $NC$}
				\IF{$nc$ inherits $api_h$}
					\STATE$R$.put($nc$)
				\ENDIF
			\ENDFOR
		\ENDIF
	\ENDFOR
  \end{algorithmic}
\end{algorithm}

First of all, Algorithm~\ref{algo:localizationB} requires three input data: 
(1) $crash$, the crash under study;
(2) \emph{ST}, which is the list of developer methods contained in the stack trace, e.g. lines 6-7 in Figure~\ref{fig:sandwich};
(3) $api_h$, the wrongly handled API, which is approximated as the associated wrongly handled API of the most similar crash present in Category B of our empirical dataset. 
More formally, let be $Crash_B$ the set of all the crashes in Category B. 
We identify the most similar crash $crash_{sim}$ by following Equation~\ref{eq:maxB}. 
Since their crash reasons are the most similar, it is with the highest possibility that both have wrongly handled the same API.

\begin{equation}\label{eq:maxB}
Sim_{crash,crash_{sim}} = max(Sim_{crash,crash_b}), crash_b \in Crash_B
\end{equation}

The algorithm starts with retrieving a set of developer methods $S$ from the entire apk that has invoked the $api_h$ (line 1).
The outmost for-loop (lines 2-19) loops over each stack frame $sf$ in the stack trace $ST$. Then based on the type of the $api_h$, there are two sub-routines: (a) when $api_h$ type is ``call-in'' (lines 4-11); (b) when $api_h$ type is ``callback'' (lines 13-17). Next we discuss both sub-routines in detail.

Sub-routine for type ``call-in'' is a for-loop (lines 4-11) that loops over each method $s$ in $S$. We then loop over (lines 5-9) all \emph{Active Methods} ($AM$) declared in the same class as $sf$, where \emph{Active Methods} are methods having actual code snippets in the Java class files, not including the inherited ones. The function \emph{links} (line 6) checks 3 sub-conditions: (1) if $s$ is invoked by $am$, or (2) if $s$ and $am$ are declared in the same Java class or (3) if an instance of the declaring class of $s$ has been passed to $am$ as a parameter. Sub-condition (1) checks if $s$ is in the extended call graph of $am$, same as locators like~\cite{wu2014crashlocator}. Sub-condition (2) implies that $s$ is a callback method that involves controling the component lifecycle as $am$ does. Sub-condition (3) implies potential data flow between $s$ and $am$. When the condition holds true in line 6, a score is added for $s$ (line 7). Here $d$ is the distance between $sf$ and \emph{Crashed API} in the stack trace. It reflects on the same intuition in Section~\ref{subsubsec:category_A}.

Sub-routine for type ``callback'' is implemented with a for-loop (lines 13-17) that loops over all the inherited Non-overridden Callback ($NC$) of the class where $sf$ is declared. If $nc$ inherits from $api_h$ (line 14), it implies that overriding it may fix the problem, therefore $nc$ will be added to the rank $R$ (line 15). With the same intuition in Section~\ref{subsubsec:category_A}, this sub-routine is designed so that when $sf$ is closer to \emph{Crashed API} in the stack trace, $nc$ is in the higher location in the rank.

Algorithm~\ref{algo:localizationB} addresses the concerns in the empirical study (cf. Section~\ref{subsec:categoryB_empricial}). It can further locate faulty methods that are not in the extended call graphs, or even methods without actual code snippets.

\subsubsection{Category C: Out of Stack Trace, out of Code}
\begin{figure}[!h]
	\centering
	\includegraphics[width=0.8\linewidth]{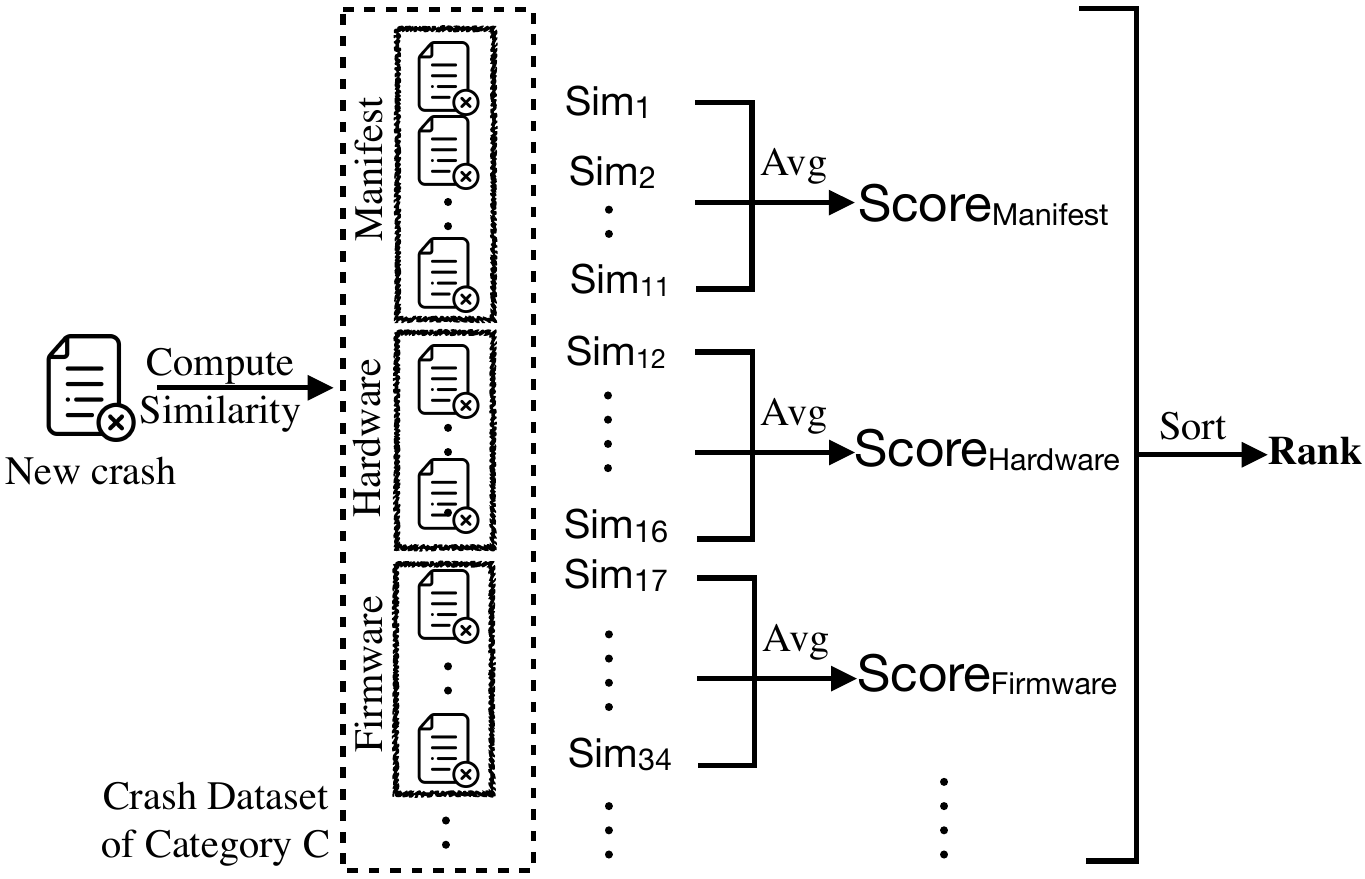}
	\caption{Localization Process for Category C.}
	\label{fig:categoryCAlgt}
\end{figure}

Figure~\ref{fig:categoryCAlgt} describes the localization process for crashes that have been classified into Category C. To infer a ranked list of potentially buggy locations, this process computes a suspiciousness score for each location.
Since the true fault locations in Category C are not in the code, the locations in this ranked list are sub-categories (e.g. manifest, asset, etc.). 

With the new crash, we start by computing the similarity score $Sim_{crash,crash_c}, crash_c \in Crash_C$. Here $Crash_C$ is the set of all the crashes of Category C in the empirical dataset. In Figure~\ref{fig:categoryCAlgt}, the similarity scores are denoted as $Sim_{caseID}$ for short. We then take an average of $Sim_{caseID}$ over the same sub-categories. Sub-categories with higher similarity scores take higher positions in the $Rank$.

\section{Experimental Setup}\label{sec:setup}
This section clarifies the research questions, the metrics used to assess \tool, and the parameter values involved.%

\subsection{Research questions}\label{sec:rqs}
We empirically validate the performance of \tool by investigating the following research questions:

\begin{itemize}
\item \textbf{RQ1:} To what extent is the categorization strategy effective?
\item \textbf{RQ2:} To what extent are the localization algorithms reliable?
\item \textbf{RQ3:} What is the overall performance of \tool{}?
\item \textbf{RQ4:} How does \tool{} perform on crashes in the wild?
\end{itemize}

\subsection{Metrics}\label{subsec:metric}
Crash localization is a recommendation problem. To measure the performance of \tool, we rely on rank-aware metrics, which are widely used in information retrieval communities and have been previously used to evaluate crash localization techniques~\cite{wu2014crashlocator}. 

{\bf Recall@k}: The percentage of crash cases whose buggy functions appear in top $k$ locations. A higher score indicates better performance of \tool{}.

{\bf MRR} (Mean Reciprocal Rank): The mean of the multiplicative inverse of the rank of the first correct location.
As defined in Equation~\ref{eq:mrr}, $Rank_i$ is the rank for the $i^{th}$ crash case, in a set of crash cases $E$. A high value of MRR means developers on average need to examine fewer locations in the rank, and therefore, a better performance~\cite{shi2012climf}.
\begin{equation}\label{eq:mrr}
MRR = \frac{1}{|E|}\sum^{|E|}_{i=1} \frac{1}{Rank_i}
\end{equation}

\subsection{Cross-validation}
We perform 5-fold cross-validation over the empirical dataset of $500$ sample crashes. 
The dataset is randomly divided into 5 subsets of $100$ sample crashes: 5 experiments are then carried where every time a specific subset of $100$ is used as ``test'' data while the remaining subsets containing the rest $400$ cases are merged to form ``training'' dataset. 
The computed performance metrics are then summed over the 5 folds.

\subsection{Feature Selection}\label{subsec:feature}
\begin{figure}[!h]
	\centering
	\includegraphics[width=0.6\linewidth]{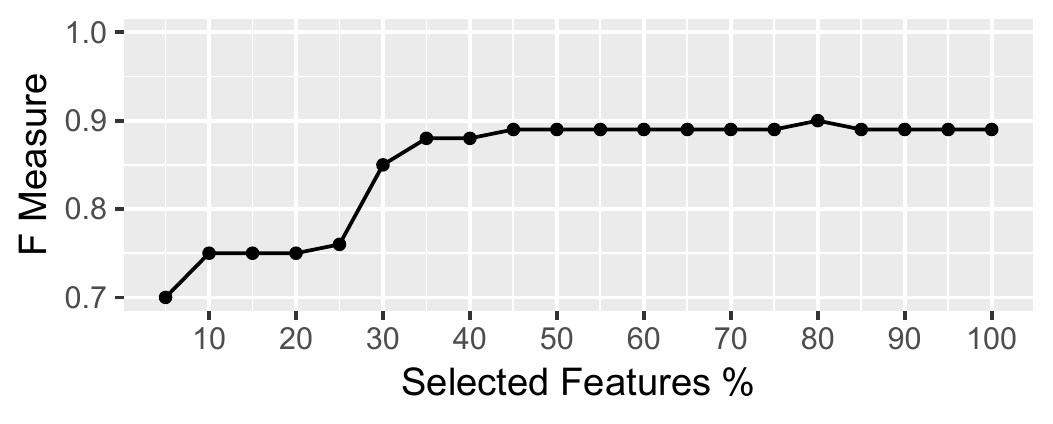}
	\caption{F Measure v.s. Selected Features.}
	\label{fig:performance_phase1}%
\end{figure}

In the empirical dataset, the vocabulary contains $1108$ unique words. 
To avoid over-fitting, we select only a portion of them for Phase 1. 
We use the $\chi^2$ test for each word~\cite{miller1982maximally}. 
A higher value of $\chi^2$ indicates a stronger correlation between the word and the category. 
Figure~\ref{fig:performance_phase1} shows the relation between the F Measure of Phase 1 and the percentage of words chosen (ranked in descending order by $\chi^2$ values).
We can see that with the top 50\% of the features, the overall performance already stabilizes. We then always use top 50\% of the words in the vocabulary.

\section{Experimental Results}\label{sec:evaluation}

\subsection{RQ1: Effectiveness of Categorization}\label{subsec:rq1}
We use our ground truth of 500 crashes to assess the performance of \tool in the first phase of the approach, namely the categorization. 
We provide in Table~\ref{tab:cross_validation_matrix} the confusion matrix as well as the precision and recall of our experimental results.
\tool{} yields a very high precision for predicting crashes in Category A, reaching 0.96. The precision for crashes in Categories B and C are comparably lower, at 0.65 and 0.60, respectively. In terms of recall, the approach is effective for Category A (0.91), Category B (0.82), and Category C (0.75). 
Overall, \tool is successful in categorizing 444 out of 500 crash samples (88.8\%).

\begin{table}[!htb]
    \caption{Effectiveness of Categorization (Phase 1)}
    \label{tab:cross_validation_matrix}
    \begin{subtable}{.6\linewidth}
      \centering
        \resizebox{1.0\linewidth}{!}{
\begin{tabular}{c |c | c |c | c}
             & \multicolumn{3}{|c|}{Actual}\\
\cline{2-5}
 & A & B & C & Total \\
\hline
Predicted as Category A &374&6&8&388\\
Predicted as Category B &20&40&2&62\\
Predicted as Category C &17&3&30&50\\
\hline
Total &411&49&40&500\\
\hline
\end{tabular}
}
    \end{subtable}%
    \begin{subtable}{.4\linewidth}
      \centering
\resizebox{1.0\linewidth}{!}{
\begin{tabular}{c|c|c}
& Precision & Recall\\
\hline
Category A & 0.96 & 0.91\\
Category B & 0.65 & 0.82\\
Category C & 0.60 & 0.75\\
\hline
\end{tabular}
}
    \end{subtable} 
\end{table}

\begin{tcolorbox}[breakable,boxsep=5pt,arc=0pt,top=0mm,left=0mm, right=0mm, bottom=0mm]
{\bf Answer to RQ1}: \tool{} is overall effective in categorizing new crash samples. However, there is still room of improving the precision when predicting samples in Categories B and C.
\end{tcolorbox}

\subsection{RQ2: Effectiveness of Localization}
To evaluate the localization phase of \tool, we consider sample crashes for each category and assess the rank localization yielded by the specific algorithm developed for that category. 
Table~\ref{tab:category_perfect} summarizes the Recall@k (with $k\in\{1,5,10\}$ and MRR.

To make sure the evaluation of Phase 2 is not affected by the outcome of Phase 1, we propose to assess the performance of localization with the assumption of perfect categorization.

\begin{table}[!h]
\centering
\caption{Localization Performance}
\label{tab:category_perfect}
\resizebox{0.9\linewidth}{!}{
\begin{tabular}{c |c | c |c |c}
\hline
Category & Recall@1 & Recall@5 & Recall@10 & MRR \\
\hline
A & 0.97(400/411) & 0.99(406/411) & 0.99(407/411) & 0.98\\
B & 0.39(19/49) & 0.61(30/49) & 0.63(31/49) & 0.48\\
C & 0.78(31/40) & 1.00(40/40) & 1.00(40/40) & 0.85\\
\hline
Total & 0.90(450/500) & 0.95(476/500) & 0.96(478/500) & 0.92\\
\hline
\end{tabular}
}
\end{table}

For cases in Category A, the true fault location can almost always be found at the top of the rank. The high value of MRR at 0.98 confirms the intuition in Section~\ref{subsubsec:category_A} that it takes much less effort in finding fault location for Category A.
For cases in Category B, the recall@1 starts at 0.39 and increased substantially for recall@5 at 0.61. One more case is successfully located with recall@10 at 0.63. The overall MRR is 0.48. Given the fact that the search space is vast (there can be tens of thousands of developer methods in the apk), Algorithm~\ref{algo:localizationB} demonstrates decent performance.
For most cases in Category C, the true sub-category of the fault location can be found topmost, with the MRR at 0.85.

\begin{tcolorbox}[breakable,boxsep=5pt,arc=0pt,top=0mm,left=0mm, right=0mm, bottom=0mm]
{\bf Answer to RQ2}: The localization algorithms (Phase 2) of \tool are reasonably effective for suggesting the correct fault location. \tool{} shows descent performance even when challenged by the vast search space for crashes in Category B.
\end{tcolorbox}

\subsection{RQ3: Overall Performance of \tool{}}

Table~\ref{tab:cross_validation_overall} summarizes the overall performance of \tool{} combining Phase 1 and 2.
The MRR of all 3 categories slightly dropped, since some cases are miscategorized in Phase 1. Clearly, the overall performance is affected by Phase 1. However, since the two phases in \tool{} are loosely coupled, we envisage improvements of overall performance in the future when better classifiers are proposed.

\begin{table}[!h]
\centering
\caption{Overall Performance of \tool{}}
\label{tab:cross_validation_overall}
\resizebox{0.9\linewidth}{!}{
\begin{tabular}{c |c | c |c |c}
\hline
Category & Recall@1 & Recall@5 & Recall@10 & MRR \\
\hline
A & 0.90(370/411) & 0.91(373/411) & 0.91(373/411) & 0.90\\
B & 0.37(18/49) & 0.59(29/49) & 0.61(30/49) & 0.46\\
C & 0.72(29/40) & 0.75(30/40) & 0.75(30/40) & 0.73\\
\hline
Total & 0.83(417/500) & 0.86(432/500) & 0.87(433/500) & 0.85\\
\hline
\end{tabular}
}
\end{table}

\begin{tcolorbox}[breakable,boxsep=5pt,arc=0pt,top=0mm,left=0mm, right=0mm, bottom=0mm]
{\bf Answer to RQ3:} \tool{} is an effective approach for locating crashing faults when they are in/outside stack traces, even outside code. Better performance is guaranteed when categorization (Phase 1) is further improved.
\end{tcolorbox}

\subsection{RQ4: Performance in the Wild}
The heuristics based on which \tool{} is built may be biased by the empirical dataset.
To mitigate this threat, we assess the effectiveness of \tool{} with a dataset selected in the wild. We want to verify to what extent \tool{} can be generalized. We leverage the independent dataset prepared by Fan et al.~\cite{fan2018large} who thoroughly (by crawling the entire GitHub) and systematically (by applying strict criteria) collected 194 crashed apks from open-source Android repositories. 
Before evaluation, we apply the constraint rules of Section~\ref{subsec:dataset_construction}, and focus on the 69 relevant crash cases that could be identified. 
Note that this dataset contains true fault locations already verified by the app developers. 
Since the cases in the dataset are from a wide time span (2011-2017), the partition is randomly decided on normal distribution over the year of app release.

\begin{table}[!htb]
\caption{Categorization on an independent dataset.}
    \label{tab:in_the_wild_matrix}
    \begin{subtable}{.6\linewidth}
      \centering
        \resizebox{1.0\linewidth}{!}{
\begin{tabular}{c |c | c |c | c}
             & \multicolumn{3}{|c|}{Actual}\\
\cline{2-5}
 & A & B & C & Total \\
\hline
Predicted as Category A &54&1&0&55\\
Predicted as Category B &3&6&0&9\\
Predicted as Category C &1&0&4&5\\
\hline
Total &58&7&4&69\\
\hline
\end{tabular}
}
    \end{subtable}%
    \begin{subtable}{.4\linewidth}
      \centering
\resizebox{1.0\linewidth}{!}{
\begin{tabular}{c|c|c}
& Precision & Recall\\
\hline
Category A & 0.98 & 0.93\\
Category B & 0.67 & 0.86\\
Category C & 0.80 & 1.00\\
\hline
\end{tabular}
}
 \end{subtable} 
\end{table}

Table~\ref{tab:in_the_wild_matrix} shows the confusion matrix, as well as the precision and recall of Phase 1 (categorization) on this independent dataset. The precision for all categories is high, reaching 0.98 (54/55), 0.67 (6/9), and 0.80 (4/5) respectively. The recalls are also high, at 0.93 (54/58) for A, 0.86 (6/7) for B, and a perfect 1.00 (4/4) for C.

Table~\ref{tab:in_the_wild_recall} provides measures for the overall performance. To compute the similarity scores which are required to locate the bug related to crashes from  Categories B and C, we use the crash records from the empirical dataset.
The recalls and MRR in Category A remain high. As for Category B, \tool{} is able to yield recall@k values and MRR of 0.43 when suggesting fault locations. As for Category C, the total MRR is at $0.43$, suggesting more stack traces in Category C might be the key for better performance.

\begin{table}[!h]
\centering
\caption{Recall@k and MRR on an independent dataset.}
\label{tab:in_the_wild_recall}
\resizebox{0.9\linewidth}{!}{
\begin{tabular}{c |c | c |c |c}
\hline
Category & Recall@1 & Recall@5 & Recall@10 & MRR \\
\hline
A & 0.72(42/58) & 0.93(54/58) & 0.93(54/58) & 0.81\\
B & 0.43(3/7) & 0.43(3/7) & 0.43(3/7) & 0.43\\
C & 0.25(1/4) & 1.00(4/4) & 1.00(4/4) & 0.40\\
\hline
Total & 0.67(46/69) & 0.88(61/69) & 0.88(61/69) & 0.74\\
\hline
\end{tabular}
}
\end{table}

\begin{tcolorbox}[breakable,boxsep=5pt,arc=0pt,top=0mm,left=0mm, right=0mm, bottom=0mm]
{\bf Answer to RQ4:} The evaluation on an independent dataset shows that \tool{} can be generalized. \tool is a milestone in this respect that it considers various crashing location cases. However, a community effort is still required to construct a representative dataset of crashes to push forward the state of the art in crashing fault localization. \end{tcolorbox}

\section{Disscussion}\label{sec:discussion}
\subsection{Comparing \tool{} with other Locators} 
Along with their empirical analysis of Android app crashes, Fan et al.~\cite{fan2018large} mentioned, in a single paragraph, a prototype crashing fault locator: ExLocator. Unfortunately, since the tool has not been publicly released, we could not directly compare it against \tool{}. We note, based on its description, however, that ExLocator has a limited usage scenario since it focuses on only 5 exception types. 
CrashLocator~\cite{wu2014crashlocator} can also locate faults outside the stack trace. However, CrashLocator needs to abstract patterns from a great number of repeated crashes of the same project. Unfortunately, for both datasets presented in this paper, this requirement is not satisfied. Moreover, CrashLocator requires source code and change tracking of the projects, unavailable for our empirical dataset. Therefore, we are not able to apply CrashLocator.

Although direct comparison in terms of effectiveness is not possible in this scenario, we can compare the applicability. \tool{} is considered to have a wider application range compared to ExLocator, i.e., it can be applied to all exception types, and considered to be more lightweight and source code independent compared to CrashLocator, i.e., it requires only the crash message and the apk.

\subsection{Developer Effort for Locating Bugs} 
In the motivating example, we demonstrated why locating buggy methods outside the stack trace can be challenging. We also want to measure the effort that developers put in locating such bugs. In Fan et al.'s dataset, each crash is documented with its duration, i.e., the time between the issue creation and its official closure by the developers. For bugs in the stack trace, it takes developers 26 days on average to close the issues. For bugs outside the stack trace, it drastically increases to 41 days. The ratio is 41/26=158\%. Although it may not always be precise to measure effort in terms of issue duration, this would confirm our observation to some extent.

\section{Threats to Validity}\label{sec:validity}

\subsection{Internal Threats}
In the empirical study presented in Section~\ref{sec:empirical}, we have manually built the ground truth of buggy locations that we made available to the community. 
Although we have tried our best to perform this manual inspection with the help of (1) the Android official documentation, (2) programmer information exchanging forums like StackOverflow or GitHub, (3) tools such as Soot or CodeInspect, there is no guarantee that all buggy locations we retrieved are the true causes for the crashes. 
This might affect the conclusions we draw from this dataset and the answers to RQ1-RQ3. 

\subsection{External Threats}
We extracted our dataset from the open benchmark ReCBench built by Kong et al~\cite{kong2019-ISSTA-crashes}. 
Although the large dataset they propose contains diverse apks collected from various popular app markets such as Google Play (ensuring a good diversity of apps), 
the collected crash cases are retrieved by testing apks with only two testing tools.  
Therefore, the yielded crashes could not be representative of the whole spectrum of crashes present in the Android ecosystem. 
Similarly, the dataset proposed by Fan et al.~\cite{fan2018large} is extracted from open source Android app GitHub repositories only.
Moreover, they have applied certain rules for collecting the crashed cases, e.g., they extract only crash bugs that have been closed by repository maintainers. 
The potential limitations with both datasets may affect the effectiveness we have shown in RQ1-RQ4.

\section{Related Work}\label{sec:related}

A recent survey by Wong et al.~\cite{wong2016survey} marks the activity of identifying the locations of faults in a program to be most tedious, time-consuming, and expensive, yet equally critical. Therefore, lots of techniques have been proposed attempting to ease the work of finding the fault locations. Although we did not find a dedicated tool for identifying locations in Android apps, there are some approaches proposed for general software programs.
For example, Wu et al. proposed CrashLocator~\cite{wu2014crashlocator} to score and rank suspicious locations that have caused program crashes. CrashLocator suggests that the buggy methods can be located in the static call graphs extended from the stack traces. However, it is not suitable to work on programs with multiple entry points and separate call graphs such as Android apps. Moreover, its scoring factors, which require source code and change histories, also limit its application scope to Android apps, for which most of them are released in a closed way (i.e., no change histories).
Gu et al.~\cite{gu2019does} proposed another approach called CraTer that adopts information retrieval techniques to predict whether the real fault resides inside the stack traces. However, CraTer is not able to suggest the actual buggy location.
BugLocator~\cite{zhou2012should} applies a revisited Vector Space Model (rSVM) to retrieve relevant files for fixing a bug on a large number of bug reports. However, its granularity falls in file level, which still requires human verification for more fine-grained location identification.
Wong et al.~\cite{wong2014boosting} build their work on top of BugLocator~\cite{zhou2012should} and leveraged stack trace to improve the approach and indeed achieved better performance.
Fan et al.~\cite{fan2018large} briefly describes a fault localization prototype ExLocator for Android apps. ExLcator only supports 5 exception types and has a limited usage scenario.
Furthermore, in the community of Automatic Program Repair (APR), statement-level fault localization is often among the first few steps. Researchers have improved it in various aspects~\cite{abreu2007accuracy,koyuncu2019ifixr,abreu2009practical,jones2005empirical,abreu2009spectrum,wang2015evaluating,lal2012static}.%

Many research works have been proposed to address Android app crashes in recent years.
For example, Fan et al.~\cite{fan2018large} performed a large scale analysis on framework-specific Android app crashes. They have invented the grouping techniques to group the Android app crash cases into buckets to study similar root causes based on each bucket. 
Researchers have also spent efforts attempting to automatically reproduce the reported crashes~\cite{li2017static, martin2016survey}.
Indeed, to achieve this purpose, Zhao et al. have proposed ReCDroid~\cite{zhao2019recdroid}, which applies a combination of natural language processing (NLP) and dynamic GUI exploration to reproduce given crashes.
G{\'o}mez et al.~\cite{gomez2016reproducing} proposed another approach for reproducing crashes by providing sensitive contexts. 
Moran et al.~\cite{moran2016automatically} further presented a prototype tool called CrashScope, aiming at generating an augmented crash report to automatically reproduce crashes on target devices.
Researchers have gone one step deeper to propose automated tools to automatically fix such identified crashes. 
Indeed, Tan et al.~\cite{tan2018repairing} have proposed an automatic repairing framework named Droix for crashed Android apps. Droix adopts 8 manually constructed fixing patterns on crashed Android apps to generate app mutants and suggest one that fixes the crash.
Following this work, Kong et al.~\cite{kong2019-ISSTA-crashes} present to the community an automatic fix pattern generation approach named CraftDroid for fixing apps suffering from crashes.

The special Android callback-based mechanism and its effect have drawn the attention of many researchers with the ever-booming of Android devices.
Yang et al.~\cite{yang2015static} targets the even-driven and multi-entry point issue of Android apps, and proposed a program representation that captures callback sequences by using context-sensitive static analysis of callback methods. 
Flowdroid~\cite{arzt2014flowdroid} targets at exposing privacy leakages on Android phones. It establishes a precise model of the Android lifecycle, which allows the analysis to properly handle callbacks invoked by the Android framework. 
Relda2~\cite{wu2016light} is a light-weight and precise static resource leak detection tool based on Function Call Graph (FCG) analysis, which handles the features of the callbacks defined in the Android framework. Together with other existing works like~\cite{yang2012leakminer, li2016droidra}, they all dealt with Android callback-based mechanism in various manners.
Although these works are different from ours, their approach in handling lifecycle and callback methods could be borrowed to enhance our approach towards better dealing with Category B crashes.

\section{Conclusions}\label{sec:conclusions}

In this work, we performed an empirical study.
This study shows that 37\% crash types are related to bugs that are outside the stack traces, which imposes challenges to the localization problem. 
We then proposed \tool{}, a two-phase categorization and localization tool that is able to generate a ranked list of bug locations for developers to examine.
The effectiveness of \tool{} is assessed with both this empirical dataset and an in-the-wild scenario on a third-party dataset. 
Our work brings inspiring insights into the crashing faults localization problem for Android apps and calls for more attention from both the developers and the research community.

\balance
\bibliographystyle{unsrt}%
\bibliography{sec/bibliography}
\end{document}